# Changes in the spectrum of Z Ursae Majoris during its rise through a maximum in June 2014

**David Boyd**
*Variable Star Section, British Astronomical Association, [davidboyd@orion.me.uk]*

**Abstract**

The semiregular variable Z UMa experienced a particularly deep minimum at the end of March 2014 followed by a three month rise to maximum. The rise and subsequent decline were followed with low resolution (R~1000) spectroscopy, possibly for the first time. The spectral type of Z UMa varied between M7III at minimum and M4III at maximum while the bright hydrogen emission lines reported by observers during the last century now appear relatively weak and transitory.

**The nature of Z UMa**

Z UMa was first classified as a RV Tauri star by Müller in 1925 (1). Gerasimovic, in a Harvard College Observatory Bulletin in 1928, reported that an analysis of visual observations showed Z UMa to be "of RV Tauri type but intimately connected with ordinary Mira variables" (2). The General Catalogue of Variable Stars (GCVS) (3) now classifies Z UMa as SRB, a semiregular late-type giant, with pulsation period 195.5 days and V magnitude range 6.2 to 9.4. Analysis of the observational data suggests that there are two pulsation cycles within the star which beat with each other to produce the observed light curve (4). Z UMa was AAVSO Variable Star of the Month in March 2000 (5). Examination of the BAAVSS database (6) shows that in recent years its visual magnitude has ranged between 6.0 and 10.0. The complexity of the internal state of variables such as Z UMa means they are still the subject of research.

**Previous spectral observations**

Z UMa first attracted attention because of its peculiar spectrum which was noted by King at the Harvard College Observatory in 1904. It was subsequently found to be variable by at least 1.5 magnitudes on 22 plates taken between 1897 and 1904. The spectrum was reported to be of Father Angelo Secchi's Type III with bright H$\delta$ and H$\gamma$ emission lines and classified according to Annie Cannon's Harvard Draper scheme as Md indicating the presence of hydrogen emission lines (7).

Working at the University of Michigan and Mount Wilson Observatories in the decade before 1923, Merrill obtained spectrograms of 117 stars, including Z UMa, which had previously been classified as Md and were brighter than mag 9.0 at maximum (8). As the plates he was using were most sensitive in the range 3800 – 5000Å, the spectra of red stars were difficult to record and required long exposures, so where possible he observed them close to maximum brightness. He noted the presence of H$\gamma$ and H$\delta$ emission lines in the spectrum of Z UMa but commented that "the bright hydrogen lines are weak" compared to the same lines in the spectra of other M-type stars he observed. Nevertheless, Merrill was able to measure the radial velocity of Z UMa using the H emission lines in its spectra. Using the extensions of the Draper scheme adopted by the IAU in 1922, he estimated the spectral type of Z UMa at maximum as M6e (the letter e indicating the presence of emission lines).

In 1942 Keenan published the results of a photographic survey of spectral types of 67 red variables at McDonald Observatory which listed Z UMa as M5III using the new MK classification system to assign an additional luminosity class III to the spectral type (9). It appears that he was primarily interested in determining the luminosity classes of these stars as he did not record the presence of emission lines for any of the stars.

The GCVS lists the spectral type of Z UMa as M5IIIe, while Simbad (10) lists it as M5IIIv (the letter v indicating intrinsic variability in the spectrum). Despite several references in the literature to the presence of bright emission lines in the spectrum of Z UMa, I have not so far found any published spectra showing these lines in order to qualitatively assess their strength.

**New observations**

On 2014 March 30, John Toone reported in a baavss-alert email (11) that Z UMa was then at an unusually deep minimum at mv=10.0. I recorded my first spectrum of Z UMa on April 1 and over the following 4 months I recorded 9 spectra as the star brightened to a maximum at mv=6.7 on June 29 before starting to fade.

The equipment used was a Celestron 280mm SCT with a LISA spectrograph and SXVR-H694 CCD camera. The spectra have a resolution of 7Å and a SNR of 350 or greater at 6000Å. All spectra were wavelength calibrated using an internal neon lamp plus Balmer absorption lines in the spectrum of the A2V type star HD 106591 located close to Z UMa, and were corrected for instrument and atmospheric extinction effects before being flux calibrated using the method I described on the ARAS Forum (12). This requires knowledge of the V magnitude of the star at the time each spectrum is recorded.

As Z UMa rose from minimum I was able to make three V magnitude measurements before the star become too bright for me to measure. I subsequently obtained a further two V magnitudes from Arne Henden taken with the AAVSO Bright Star Monitor (13) and twelve V magnitudes from the AAVSO database (14). These latter values were mostly from the same observer so should be internally consistent. As the star faded, no V magnitude measurements were available from the AAVSO and it was still too bright for me to obtain photometry.

I investigated the relationship between the V magnitudes measured as the star brightened and the mv magnitude estimates made over the same period by Toone (6). Both the V and mv magnitudes were well fitted by second order polynomials of the Julian Date (JD) as shown in Figure 1. I found that the relationship between V and mv magnitudes could be well fitted by the linear equation

$$V = mv*0.8456 + 0.9350 \qquad (1)$$

Using the second order polynomial for V magnitudes, I computed V magnitudes at the times of each of my spectra as the star brightened. By fitting a straight line to Toone's mv estimates after maximum and using equation (1), I computed V magnitudes at the times of each of my spectra as the star faded. These computed V magnitudes were used to flux calibrate each spectrum. Table 1 lists these V magnitudes for each date and JD on which I recorded a spectrum. These dates are also indicated by arrows in Figure 1.

Total galactic visual extinction E(B-V) in the direction of Z UMa is 0.029 (15). Extinction to the light from Z UMa will be less and thus will produce a barely detectable effect on these spectra so no correction for extinction has been made.

A composite plot of the flux-calibrated spectra obtained as Z UMa brightened is shown in Figure 2 and as it started to fade in Figure 3. The spectra are dominated by strong absorption bands of the diatomic titanium oxide molecule, TiO. These are characteristic of spectral class M (Morgan, Keenan & Kellman (16)) and indicate that the temperature of the photosphere is less than about 3500K. Matching these spectra against spectra from the Pickles Stellar Spectral Flux Library (17) using the winmk software (18), the spectral type changed from between M6III and M7III at minimum to between M4III and M5III at maximum. Weak H$\delta$ and H$\gamma$ emission lines are visible in the spectrum at maximum but fade quickly afterwards. Given earlier references to bright emission lines, albeit weak ones, without any published spectra from that era it is difficult to know whether these features are now relatively weaker and shorter lived than they were in the past.

**Acknowledgements**

I am very grateful to John Toone for alerting me to this opportunity to monitor the spectral changes in Z UMa as it brightened from a deep minimum, and to John Toone and Robin Leadbeater for their helpful suggestions for improving the paper. I thank Arne Henden for supplying V magnitudes from the AAVSO Bright Star Monitor. I acknowledge the BAAVSS database as a source of data on which this article is based and also the variable star observations from the AAVSO International Database contributed by observers worldwide and used in this research.

| Date | JD | V mag |
|---|---|---|
| 01-Apr-14 | 2456749.45270 | 9.30 |
| 14-Apr-14 | 2456762.44710 | 9.00 |
| 04-May-14 | 2456782.42690 | 8.49 |
| 25-May-14 | 2456803.45160 | 7.86 |
| 25-Jun-14 | 2456834.44320 | 6.79 |
| 09-Jul-14 | 2456845.45764 | 6.69 |
| 13-Jul-14 | 2456855.47708 | 6.90 |
| 21-Jul-14 | 2456863.45278 | 7.07 |
| 25-Jul-14 | 2456867.46181 | 7.16 |

Table 1: Date, JD and computed V magnitude for each spectrum recorded.

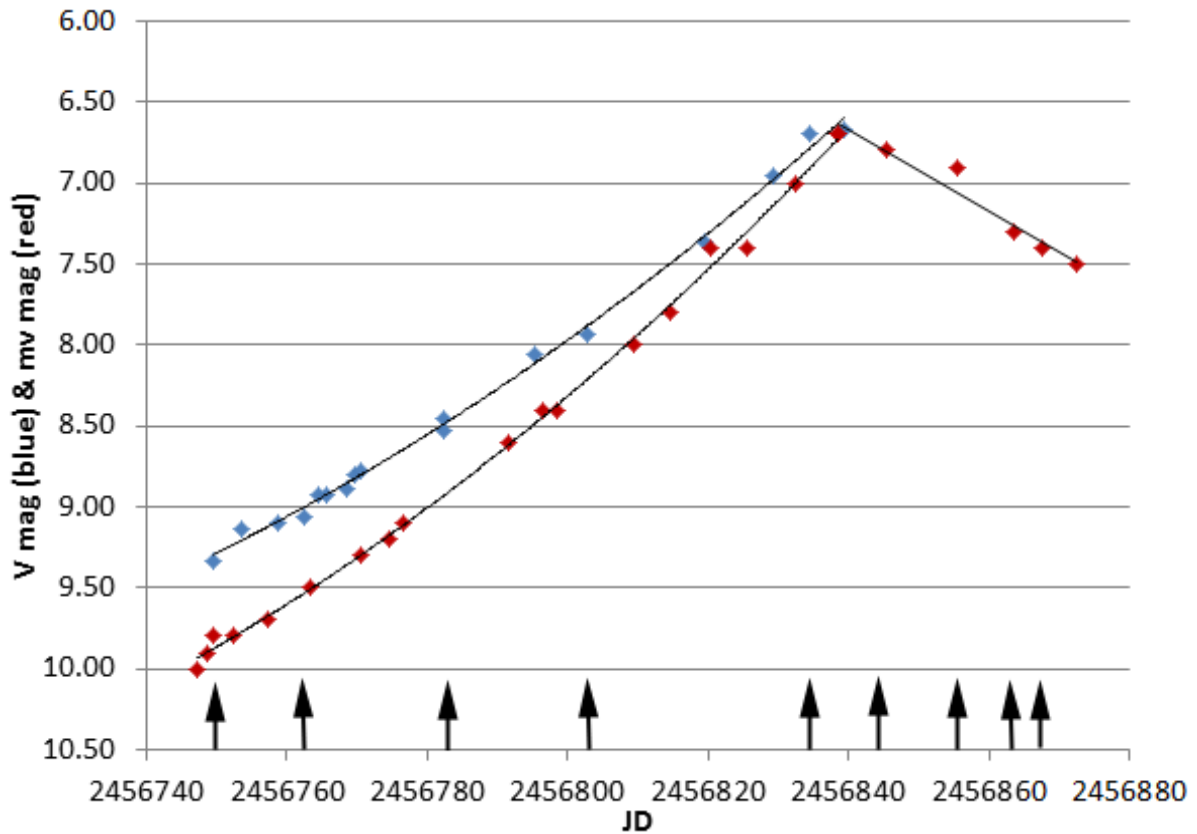

Figure 1: V magnitudes (blue) and mv estimates (red) of Z UMa with quadratic and linear fits. Arrows indicate the days on which spectra were recorded.

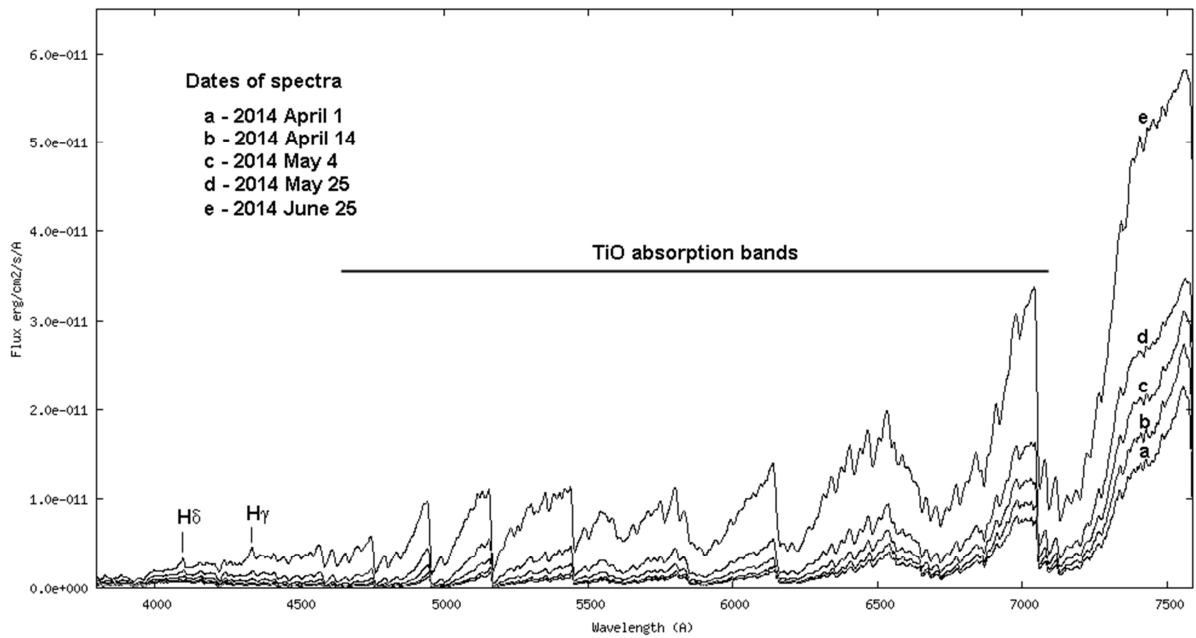

Figure 2: Flux-calibrated spectra of Z UMa as it rose from minimum to maximum.

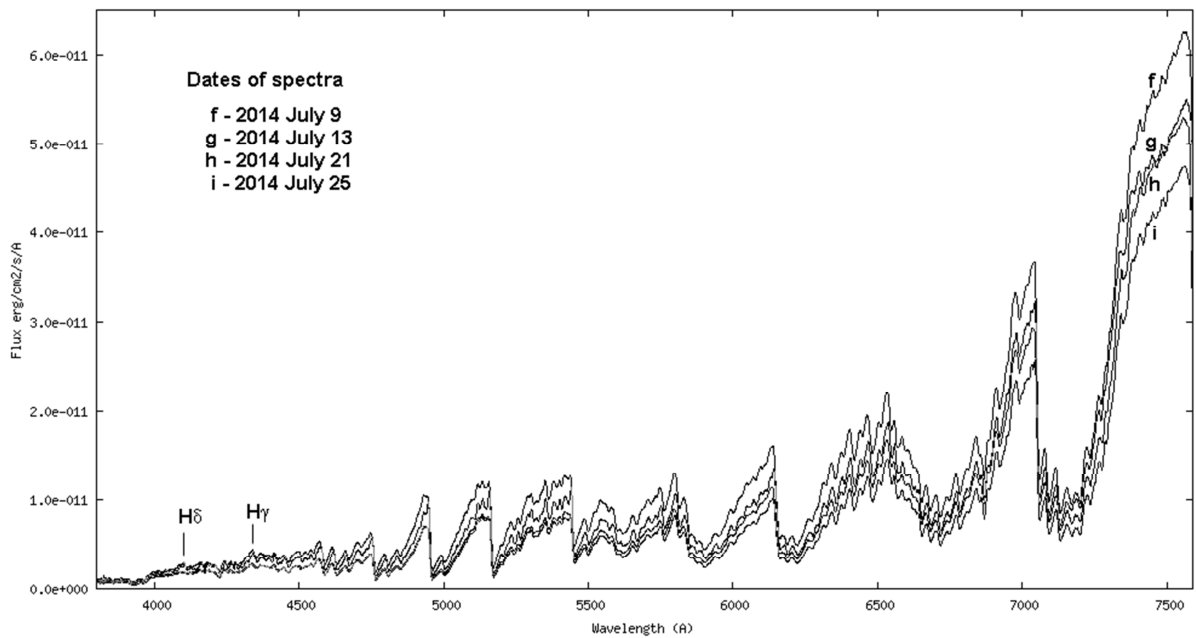

Figure 3: Flux-calibrated spectra of Z UMa as it started to decline from maximum.